\documentclass[prl,twocolumn,preprintnumbers,amsmath,amssymb,showpacs]{revtex4}

\usepackage{color}
\usepackage{graphicx}

\def\be{\begin{equation}}
\def\ee{\end{equation}}
\def\l{\lambda}
\def\Tr{{\rm Tr}}

\def\s{\sigma}

\def\a{\alpha}

\def\bea{\begin{eqnarray}}
\def\eea{\end{eqnarray}}
\begin{document}

\title{Parity effects in the scaling of block entanglement in gapless
spin chains} 
\author{Pasquale Calabrese${}^{1}$, Massimo Campostrini${}^{1}$, 
Fabian Essler${}^{2}$, and  Bernard Nienhuis${}^{3}$}

\affiliation{
$^{1}$Dipartimento di Fisica dell'Universit\`a di Pisa and INFN,
56127 Pisa, Italy\\ 
$^2$The Rudolf Peierls Centre for Theoretical Physics, Oxford
University, Oxford OX1 3NP, UK\\
$^3$Institute for Theoretical Physics, Universiteit van Amsterdam,
1018 XE Amsterdam, The Netherlands.}

\date{\today}

\begin{abstract}
We consider the R\'enyi $\a$-entropies for Luttinger liquids (LL). For 
large block lengths $\ell$ these are known to grow like $\ln\ell$.
We show that there are subleading terms that oscillate with frequency
$2k_F$ (the Fermi wave number of the LL) and exhibit a
universal power-law decay with $\ell$. The new critical exponent is
equal to $K/(2\a)$, where $K$ is the LL parameter. We
present numerical results for the anisotropic XXZ model and the full
analytic solution for the free fermion (XX) point. 
\end{abstract}

\pacs{64.70.Tg, 03.67.Mn, 75.10.Pq, 05.70.Jk}

\maketitle

\begin{figure}[b]
\includegraphics[width=.48\textwidth]{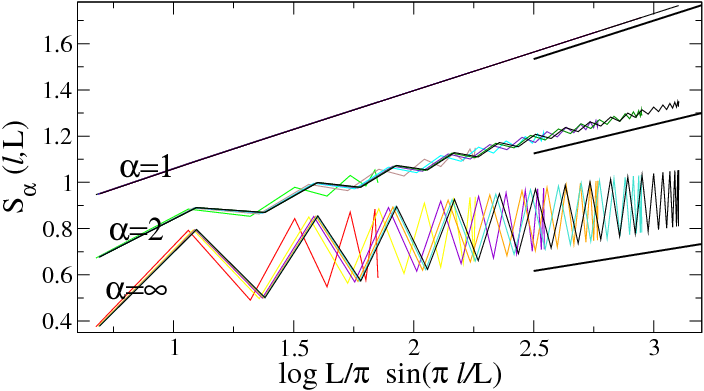}
\caption{Parity effects in R\'enyi entropies in the XXZ model at
$\Delta=-1/2$. $S_\a(\ell,L)$ for several $L$ and $\a=1,2,\infty$. 
The straight lines indicate the asymptotic slopes $(1+\a^{-1})/6$.
}
\label{2exs}
\end{figure}

Luttinger liquid (LL) theory describes the low-energy (large-distance)
physics of gapless one-dimensional models such as quantum spin chains
and correlated electron models. It corresponds to a conformal field
theory (CFT) with central charge $c=1$ and is known to provide
accurate predictions for universal properties of many physical
systems. LL theory has been applied successfully to recent experiments
on carbon nanotubes \cite{cn}, spin chains \cite{sc}, and cold atomic
gases \cite{cg}. A much studied example of a lattice model
that gives rise to a LL description at low energies is the spin-1/2
Heisenberg XXZ chain
\be
H=-\sum_{j=1}^L [\s^x_j\s^x_{j+1}+\s^y_j \s^y_{j+1}+\Delta
\s^z_j\s^z_{j+1}]\, . 
\label{H0.5}
\ee
Here $\s_j$ are Pauli matrices at site $j$ and we have imposed
periodic boundary conditions. Recent years have witnessed a
significant effort to quantify the degree of entanglement in many-body
systems (see e.g. \cite{rev} for reviews). Among the various measures,
the entanglement entropy (EE) {has been by far} the most studied. 
By partitioning an extended quantum system into two subsystems,
the EE is defined as the von Neumann entropy of the reduced density
matrix $\rho_A$ of one of the subsystems. The leading contribution to
the EE of a single, large block of length $\ell$ can be derived by
general CFT methods \cite{Holzhey,Vidal,cc-04}. The case of a
subsystem consisting of multiple blocks requires a model dependent
treatment, but the EE can still be obtained from CFT \cite{fps-08}.  
On the other hand, little is known with regard to corrections to
the leading asymptotic behaviour. In the following we consider the
R\'enyi entropies  
\be
S_\a=\frac1{1-\a}\ln \Tr \rho_A^\a\,,
\ee
which give the full spectrum of $\rho_A$ \cite{cl-08} and are
fundamental for understanding the scaling of algorithms based on
matrix product states \cite{s-07,t-08,pmtm-08}. 
We note that $S_1$ is  the von Neumann entropy and $S_\infty$ gives
minus the logarithm of the maximum eigenvalue of the reduced density
matrix (known as single copy entanglement \cite{singcopy,olec-06}). 
According to CFT, in an infinite gapless one-dimensional model
a block of length $\ell$  has entropies \cite{Holzhey,Vidal,cc-04}
\be
S_\a^{\rm CFT}(\ell)=\frac{c}6 \left(1+\frac1\a\right) \ln \ell + c'_\a\,,
\label{CFTL}
\ee
where $c$ is the central charge and $c'_\a$ a non-universal constant.
In a finite system of length $L$,  the block length $\ell$ 
in (\ref{CFTL}) should be replaced with the chord distance $D(\ell,L)=
\frac{L}\pi \sin \frac{\pi \ell}L $.   
In many lattice models the asymptotic scaling is obscured by large
oscillations proportional to $(-1)^\ell$.  
Some typical examples are shown in Fig.~\ref{2exs}, where we plot
$S_\a(\ell,L)$ for $\a=1,2,\infty$ for the XXZ model at $\Delta=-1/2$
as obtained by density matrix renormalization group (DMRG) computations.  
While $S_1$ is smooth, $S_{\a\neq1}$ is seen to exhibit large
oscillations. For $\a=\infty$ in particular it is difficult to
recognize the CFT scaling behaviour (\ref{CFTL}). While such oscillations have
been observed in several examples \cite{lsca-06,osc} and can be seen
to arise from strong antiferromagnetic correlations, a quantitative
understanding of these features was until now lacking. 
We show that  these oscillations 
obey the universal scaling law
\be
S_\a(\ell)-S_\a^{\rm CFT}(\ell)=f_\a \cos(2k_F\ell) |2\ell \sin k_F|
^{-p_\a} \,, 
\label{Sconj}
\ee
where $p_\a$ is a universal critical exponent equal to $2K/\a$. Here
$K$ the LL parameter, $k_F$ is the Fermi momentum, and $f_\a$ is a
non-universal constant. In a finite system, the block length $\ell$ in
(\ref{Sconj}) is replaced by the chord distance, and $f_\a$ is
multiplied by a {\it universal} scaling function $F_\a(\ell/L)$, that
in general depends on the parity of $L$. We note that in zero magnetic field
(half-filling) we have $k_F=\pi/2$ and the oscillating factor in
(\ref{Sconj}) reduces to $(-1)^\ell$ as observed.  
While we establish (\ref{Sconj}) for the particular case of the
Heisenberg XXZ chain (\ref{H0.5}), where the LL parameter is given by
$K={\pi}/({2 \arccos \Delta})$, we expect the scaling form to be universal, because it is related to the low-energy excitations of the model and is therefore encoded in the continuum LL field theory 
description.
Recent results for the entanglement entropy confirm these expectations \cite{rec}.

{\it XX model}. This case corresponds to $\Delta=0$ in
(\ref{H0.5}). The LL parameter and exponent in (\ref{Sconj}) are $K=1$
and $p_\a=2/\a$ respectively. The computation of the R\'enyi entropies
can be achieved by exploiting the Jordan-Wigner mapping to free
fermions, which reduces the problem to the
diagonalization of an $\ell \times \ell$ correlation matrix (see
\cite{pesc} for details).
Jin and Korepin {(JK)} showed
\cite{jk-04} that R\'enyi entropies can be obtained by the following
contour integral encircling the segment $[-1,1]$ of the real axis
\be
S_\a(\ell)= \frac1{2\pi i}\oint e_\a(\l) \frac{d\ln D_\ell(\l)}{d\l} d\l\,.
\label{oint}
\ee
Here $D_\ell(\l)$ is the determinant of a  $\ell\times\ell$ Toeplitz matrix and 
$e_\a(\l)=\frac1{1-\a} \ln\left[\left(\frac{1+\l}2\right)^\a+\left(\frac{1-\l}2\right)^\a\right]$. 
In \cite{jk-04} the Fisher-Hartwig formula was used to determine
the asymptotic scaling of the R\'enyi entropy with $\ell$, which
agrees with the CFT formula (\ref{CFTL}). Here we employ the
{\it generalized} Fisher-Hartwig conjecture \cite{gFH} in order to
go beyond the results of \cite{jk-04} and determine the subleading
corrections. The terms in the asymptotic expansion of the determinant
relevant for calculation of the R\'enyi entropy can be cast in the form
\begin{multline}
\frac{D_\ell(\l)}{D_\ell^{JK}(\l)}=1+ e^{-2i k_F \ell} L_k^{-2(1+2\beta(\l))} 
\frac{\Gamma^2(1+\beta(\l))}{\Gamma^2(-\beta(\l)))}\\+
e^{2i k_F \ell} L_k^{-2(1-2\beta(\l))} 
\frac{\Gamma^2(1-\beta(\l))}{\Gamma^2(\beta(\l)))}\,,
\label{corrFH}
\end{multline}
where $D_\ell^{JK}$ is the result of \cite{jk-04}, $2 \pi
i\beta(x)=\ln[(1+x)/(1-x)]$, and $L_k=|2\ell\sin k_F|$. The
calculation of the integral in (\ref{oint}) is now straightforward.
One expands $\ln D_\ell$ in (\ref{corrFH}) in powers of $L_k$,
determines the discontinuity across the cut $[-1,1] $, changes the
integration variable from $\lambda$ to $-i\beta(\l)$ and finally
obtains the leading behavior from the poles closest to the real axis
(details will be reported elsewhere \cite{ce-prep}). 
The resulting asymptotic expression is valid at fixed $\a$, in the
limit $\ln L_k\gg \a$. The final result is given by Eq. (\ref{Sconj})
with  
\be
f_\a= \frac{2}{1-\a}  \frac{\Gamma^2( (1+\a^{-1})/2)}{\Gamma^2((1-\a^{-1})/2)}\,.
\ee 
We note that $f_1=0$ and {therefore no oscillating
corrections for the Von-Neumann entropy are predicted}, in agreement with numerical
observations. 

The requirement that $\ln L_k\gg \a$ implies that the asymptotic behaviour
is only reached for very large block lengths, e.g.
at $\a=10$ we need $L_k\gg 20000$. In the pre-asymptotic regime there
are several sources of corrections. First the integral is no longer
dominated by the poles closest to the real axis, which leads to power
law corrections of the form  $L_k^{-2m /\a}$ (with integer $m$), which
oscillate as $e^{\pm i 2k_F \ell}$. Corrections with different
oscillatory behaviour arise from the higher order terms in the
expansion of $\ln D_\ell(\l)$ in powers of $L_k$. The first correction
is proportional to $e^{\pm i 4k_F \ell}$, the next to $e^{\pm i 6k_F
  \ell}$ etc. In zero magnetic field, where $k_F=\pi/2$, the 
leading term is proportional to $(-)^\ell$ while the second does not
oscillate. Hence there is a subleading constant background in addition
to the leading oscillatory behaviour.
In the limit $\a\to\infty$ all these terms become of the same order,
so that we need to resum the entire series that arises from
expanding $\ln D_\ell(\l)$ and then carrying out the $\lambda$ integral.
In the zero magnetic field case we  thus obtain
\be
S_\infty(\ell)-S_\infty^{JK}(\ell)=
\begin{cases}\displaystyle
\frac{\pi^2}{12} \frac1{\ln b L_k} & \ell \;{\rm odd } \,,\\ \displaystyle
- \frac{\pi^2}{24} \frac1{\ln b L_k}& \ell \; {\rm even }\,,
\end{cases}
\label{sce}
\ee
Here the constant $b\approx 7.1$ has been fixed by summing certain
contributions to all orders in $1/(\ln L_k)$ and agrees well with
numerical results \cite{foot0}.
\begin{figure}[t]
\includegraphics[width=.48\textwidth]{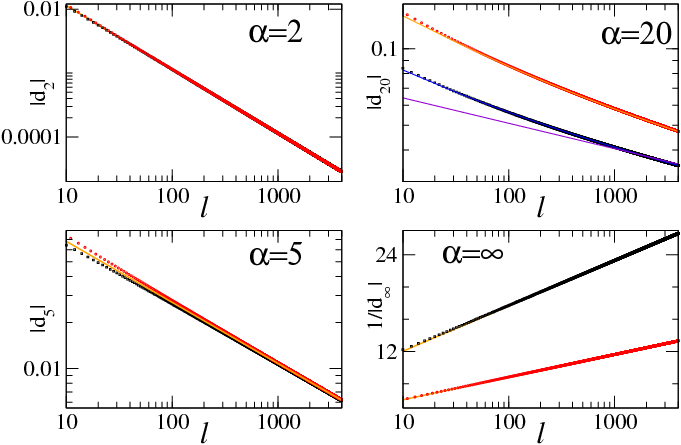}
\caption{Corrections to scaling $d_\a(\ell)= S_\a(\ell)-S_\a^{\rm CFT}(\ell)$
for the XX model and four different values of $\alpha$.  
Red circles and black squares correspond to even and odd $\ell$ respectively.
}
\label{XXTD}
\end{figure}

{\it Numerical results for the XX model} can be obtained by
diagonalizing the correlation matrix both infinite and 
finite systems. We first present the results  for infinite systems. 
We consider only the model in zero magnetic field and plot the
quantity 
\be 
d_\a(\ell)\equiv S_\a(\ell)-S_\a^{\rm CFT}(\ell)
\label{da}
\ee
where the value for the constant contribution $c'_\a$ in 
$S_\a^{\rm CFT}(\ell)$ is taken from \cite{jk-04}.
{According to Eq. (\ref{Sconj}) for $k_F=\pi/2$, $d_\a(\ell)\simeq (-)^\ell(2 \ell)^{-p_\a} f_\a$.}
In Fig.\ref{XXTD} we compare the absolute value of $d_\a(\ell)$ for
$\a=2,5,20,\infty$ and block sizes $\ell$ up to 4000 sites to our
asymptotic results (\ref{Sconj}), (\ref{sce}).
For $\a=2$ the curves for odd and even $\ell$ are practically 
indistinguishable (the line corresponding to the analytical result is 
invisible under the data points). For $\a=5$, we still obtain power laws
with the exponent $p_5=2/5$, but the curves are not as symmetrical as
for $\a=2$ {because subleading corrections become visible}.
Increasing $\a$ further, the deviations of $d_\a(\ell)$ for
$\ell<4000$ from the asymptotic behaviour become quite pronounced.
This is shown in Fig.\ref{XXTD} for the case $\a=20$, where the
leading asymptotic result (straight line) is seen to be a poor
approximation to $d_\a(\ell)$ for even $\ell$. 
Including the subleading corrections gives curves perfectly covered by data in Fig.\ref{XXTD}. 
The last panel in Fig.\ref{XXTD} shows the result in the
$\a=\infty$. The numerical results are seen to be in perfect agreement
with Eq. (\ref{sce}).

\begin{figure}[t]
\includegraphics[width=.48\textwidth]{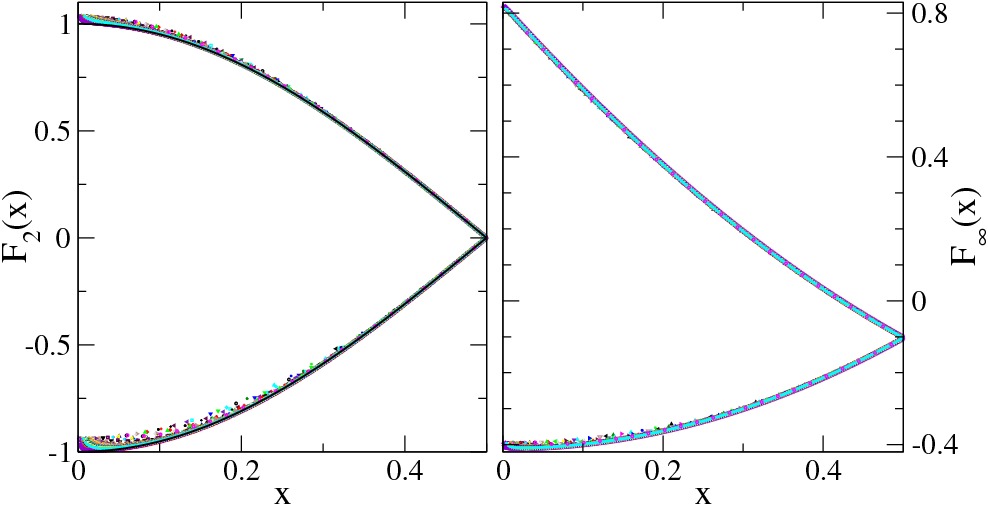}
\caption{Corrections to scaling in the finite length XX model for odd $L$. 
$F_2(x)$ and  $F_\infty(x)$ ($x=\ell/L$) obtained from 69  different systems with lengths in the range 
$17\leq L\leq 4623$ exhibit perfect data collapse.}
\label{Sscal}
\end{figure}

We now turn to finite systems. 
We numerically determined the quantity (recall that
$D(\ell,L)=\frac{L}\pi \sin \frac{\pi \ell}L $)
\be
F_\a(\ell/L)= [S_\a(\ell,L)-S_\a^{CFT}(\ell,L)] f_\a^{-1} D(\ell,L)^{2/\a}\,,
\ee
(for $\a=\infty$ we multiply by $\ln b D(\ell,L)$) for a variety of
values of $\alpha$ and system 
sizes ranging from $L=17$ to $L=4623$.
We observe that there is data collapse for any $L$ on two scaling
functions for $\ell$ odd and even respectively. Results for the
cases $\a=2,\infty$ and odd $L$ are shown in Fig.\ref{Sscal}. 
The quality of the collapse is impressive considering that 
there are no adjustable parameters and that the plots contain millions
of points ranging over three orders of magnitude in both $\ell$ and $L$. 
For $\a=2$ we observe that $F_2(x)=\pm\cos\pi x$ (these are shown as
continuous curves in Fig.\ref{Sscal}). We currently have no analytical
derivation of this scaling function. For other values of $\a$
we obtain similar data collapse, but the quality decreases with
increasing $\a$, indicating the presence of other corrections. For
even $L$ we obtain different scaling functions -- $F_\a(x)$ then
is almost constant (see below).

\begin{figure}[t]
\includegraphics[width=.45\textwidth]{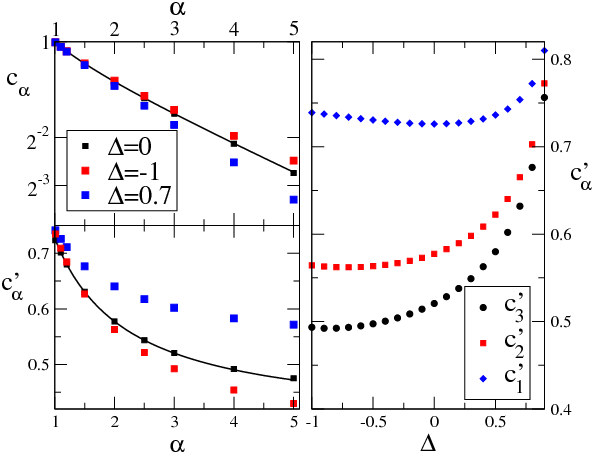}
\caption{Left: The additive constant in the R\'enyi entropies $c_\a'$ (bottom) and
  the  multiplicative constant  $c_\a$ for the moments $\Tr\rho_A^\a$ (top in log-scale) 
  as function of $\a$  at fixed $\Delta$. The drawn lines are the exact values
  for $\Delta=0$. Right: $c_\a'$ as function of $\Delta$ at fixed $\a$.}
\label{c_alpha}
\end{figure}

{\it XXZ model and DMRG}.
To characterize the XXZ model in the gapless phase with $-1\leq\Delta<1$ we
performed extensive DMRG calculations at finite $L$. 
We used the finite-volume algorithm keeping $\chi=800$ states in the
decimation procedure.  
This rather large value of $\chi$ is required to obtain a precise 
determination of the full spectrum of the reduced density matrix in
the case of periodic boundary conditions.
The data we have used in our analysis can be considered as numerically
exact. Hence the main limitation as compared to the XX case is the
relatively small value of $L$ accessible by DMRG (we considered $21 \leq L\leq 81$ for odd $L$
and $20\leq L\leq 80$ for even $L$). Another complication
stems from the fact that the value of the constant contribution
$c'_\a$ to the R\'enyi entropy is not known analytically for
$\Delta\neq0$\cite{foot2}. We obtain it by fitting our numerical
data. The results are shown in Fig. \ref{c_alpha}.
The data for $\Delta=0$ is in good agreement with the exact results of
\cite{jk-04} (full line), establishing the correctness of our fitting
procedure and the reliability of DMRG. 
The multiplicative constant $c_\a=e^{(1-\a)c'_\a}$ in the moments of the density
matrix $\Tr\rho_A^\a=c_a\ell^{c/6(\a-1/\a)}$ in Fig. \ref{c_alpha}  shows an exponential decay 
with $\a$, except for $\alpha$ very close to 1.
Hence $c_\a$ can essentially be absorbed into a rescaled block length $\ell$ as was pointed 
out in \cite{cl-08,pm-09}.  

\begin{figure}[b]
\includegraphics[width=0.48\textwidth]{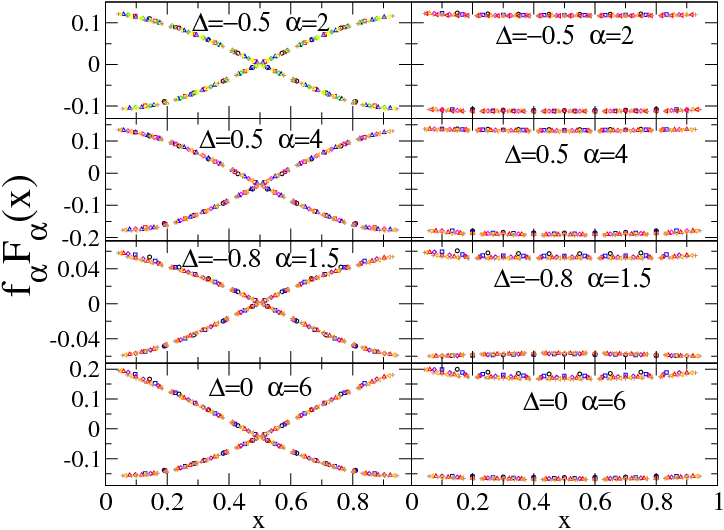}
\caption{Universal scaling function $f_\a F_\a(x)$ (where $x=\ell/L$) for
the XXZ chain for several values of $\alpha$ and $\Delta$.
Left: Several $\a$ and $\Delta$ for odd $L$.
Right: even $L$. In the latter case $F_\a(x)$ is practically 
independent on $x$.}
\label{Fig1dmrg}
\end{figure}
Having obtained the constant contribution $c'_\a$ one can determine
the universal scaling functions $F_\a(x)$. We present results for both
even and odd $L$ and a number of representative values of $\alpha$ and
$\Delta$ in Fig.\ref{Fig1dmrg}. We find that the data collapse is very
good for all cases. This is remarkable given the limited system sizes
accessible by DMRG. We note that, as expected, the data
collapse becomes poor in the vicinity of the two isotropic points
$\Delta=\pm 1$. At $\Delta=-1$ there is a marginal operator 
(see e.g. \cite{lsca-06}) that gives rise to well known logarithmic
corrections to scaling for correlation functions. 
In the ferromagnetic limit $\Delta\to 1$ the model loses conformal
invariance (the dispersion relation becomes quadratic) and is no
longer described by a LL. Hence none of the results presented here is
expected to hold.

{\it The critical Ising chain} has no strong antiferromagnetic
correlations and we therefore expect the corrections to scaling to be
non-oscillatory. It is easy to see that this is indeed the case.
Igloi and Juhasz \cite{ij-08} showed that the EEs of the XY  chain can
be expressed in terms of the EEs of two Ising  chains. At the quantum
critical point this relation reads $S^{XX}_\a(2\ell,2L)= 2S^{\rm I}_\a
(\ell,L)$, where $S^{\rm I}_\a$ refers to the critical Ising  chain.
This implies that the R\'enyi entropies in the Ising chain are
just one half of the corresponding entropies in a XX chain of twice
the block length and twice the system size. 
As both $\ell$ and $L$ are even, our results for the XX model
imply that the corrections to scaling are non-oscillatory and decay as
$\ell^{-2/\a}$. This agrees with numerical computations.

{\it Conclusions}. In this letter we considered the R\'enyi entropies
for the critical spin-1/2 Heisenberg XXZ chain with periodic boundary
conditions. 
 By a combination of analytic and numerical techniques, we computed oscillating
corrections to scaling which are expected to be universal.
These are parametrized in terms of the Luttinger parameter $K$ and the Fermi
momentum $k_F$. We argued that our results hold generally for
Luttinger liquids. In the case of open boundary conditions, a similar
relationship is expected to hold, but with the replacement $K\to K/2$
(see also \cite{lsca-06}). It would be interesting to prove this
at least in the special case of free fermions. Finally we would like
to comment on our results in light of a recent proposal \cite{fps-08},
that one way of distinguishing between different theories with the
same central charge is to consider the entanglement of multiple
blocks. Our results establish that it is sufficient to consider a
single block once one takes into account the universal subleading
oscillatory corrections. 

{\it Acknowledgments}. 
We thank J. Cardy for very helpful discussions. This work was
supported by the ESF network INSTANS (PC) and the EPSRC under grant
EP/D050952/1 (FHLE).


\begin{thebibliography}{99}

\bibitem{cn}
H. Ishii et al., Nature {\bf 426}, 540 (2003).

\bibitem{sc}
M. Klanjsek et al., Phys. Rev. Lett. {\bf 101}, 137207 (2008);
B. Thielemann et al., ibid {\bf 102}, 107204 (2009).

\bibitem{cg}
B. Paredes et al., Nature {\bf 429}, 277 (2004);
T. Kinoshita et al., Science {\bf 305}, 1125 (2004);
A. H. van Amerongen et al., Phys. Rev. Lett. {\bf 100}, 090402 (2008).

\bibitem{rev}
L. Amico et al.,
Rev. Mod. Phys, {\bf 80}, 517 (2008); 
J. Eisert et al., 
ibid. {\bf 82}, 277 (2010);
P.~Calabrese, J.~Cardy, and B. Doyon, J. Phys. A {\bf 42}, 500301 (2009).


\bibitem{Holzhey} C. Holzhey et al.,
Nucl. Phys. B {\bf 424}, 443 (1994).

\bibitem{Vidal}
G. Vidal et al., 
Phys. Rev. Lett. {\bf 90}, 227902 (2003);
J. I. Latorre et al., 
Quantum Inf. Comput. {\bf 4}, 048 (2004).

\bibitem{cc-04} P.~Calabrese and J.~Cardy, J. Stat. Mech. P06002 (2004);
J. Phys. A  {\bf 42}, 504005 (2009).

\bibitem{fps-08}
S. Furukawa et al., Phys. Rev. Lett. {\bf 102}, 170602 (2009);
P.~Calabrese et al., J. Stat. Mech P11001 (2009);
V. Alba et al., Phys. Rev. B to appear 0910.0706.

\bibitem{cl-08}
P. Calabrese, A. Lefevre, Phys. Rev. A {\bf 78}, 032329 (2008).

\bibitem{s-07} 
N. Schuch et al., Phys. Rev. Lett. {\bf 100}, 030504 (2008);
D. Perez-Garcia et al., 
Quantum Inf. Comput. {\bf 7}, 401 (2007).

\bibitem{t-08}
L. Tagliacozzo et al., 
Phys. Rev. B {\bf 78}, 024410 (2008).

\bibitem{pmtm-08}
F. Pollmann et al, Phys. Rev. Lett. {\bf 102}, 255701 (2009).


\bibitem{singcopy}
J. Eisert and M. Cramer,
Phys. Rev. A {\bf 72}, 42112 (2005); 
I. Peschel and J. Zhao, 
J. Stat. Mech. P11002 (2005). 

\bibitem{olec-06}
R. Orus et al., 
Phys. Rev. A {\bf 73}, 060303 (2006). 


\bibitem{lsca-06}
N. Laflorencie et al.,
Phys. Rev. Lett. {\bf 96}, 100603 (2006); J. Phys. A  {\bf 42}, 504009 (2009).

\bibitem{osc} G. De Chiara et al.,
J. Stat. Mech. P03001 (2006);
A. Laeuchli and C. Kollath, ibid (2008) P05018;
B. Nienhuis et al.,  ibid (2009) P02063;
H.-Q. Zhou et al.,
Phys. Rev. A {\bf 74}, 050305 (2006);
\"O. Legeza et al., Phys. Rev. Lett. {\bf 99}, 87203 (2007);
A. B. Kallin et al., ibid. {\bf 103}, 117203 (2009);
G. Roux et al., Eur. Phys. J. B {\bf 68}, 293 (2009);
I. J. Cirac and G. Sierra, 0911.3029; J. C. Xavier 1002.0531.

\bibitem{rec}
H. F. Song et al., 1002.0825.

\bibitem{pesc}
I. Peschel and V. Eisler, J. Phys. A {\bf 42}, 504003 (2009).

\bibitem{jk-04}
B.-Q.~Jin, V.~E.~Korepin, J. Stat. Phys. {\bf 116}, 79 (2004);
V.~E.~Korepin and A.~R.~Its,  ibid. {\bf 137}, 1014 (2009).

\bibitem{gFH}
see e.g. 
E.L. Basor and K.E. Morrison, Lin. Alg and its 
Appl. {\bf 202}, 129 (1994).

\bibitem{ce-prep}
P. Calabrese and F. H. L. Essler, in preparation.

\bibitem{foot0}For $\a=\infty$, logarithmic  corrections  have been 
also derived in \cite{olec-06}, but with no oscillating part. 
The predicted prefactor $-\pi^2/6$ does not agree with Eq. (\ref{sce}).



\bibitem{pm-09} F. Pollmann and J. E. Moore, 0910.0051.


\bibitem{foot2}The calculation of 
this  constant is a challenging problem for Bethe ansatz integrable systems. 
Progresses have been reported in V. Alba et al., J. Stat. Mech.  P10020 (2009).

\bibitem{ij-08}
F. Igloi and R. Juhasz,
Europhys. Lett. {\bf 81}, 57003 (2008).





\end{thebibliography}
\end{document}